\documentclass{jpsj3}
\usepackage[dvips]{color}

\title{Superfluid state in the periodic Anderson model 
with attractive interactions}

\author{Akihisa \textsc{Koga}$^{1}$
\thanks{E-mail address: koga@phys.titech.ac.jp}
and Philipp \textsc{Werner}$^{2}$}
 
\inst{$^{1}$Department of Physics, Tokyo Institute of Technology, Tokyo 152-8551, Japan \\
$^{2}$Theoretische Physik, 
ETH Zurich, Z\"urich 8093, Switzerland}

\abst{
We investigate the periodic Anderson model with attractive interactions
by means of dynamical mean-field theory (DMFT). 
Using a continuous-time quantum Monte Carlo impurity solver, 
we study the competition between the superfluid state and
the paramagnetic Kondo insulating state, and determine the phase diagram.
At the chemical potential-induced phase transition 
from the Kondo insulating state to the superfluid state,
a low-energy peak characteristic of the superfluid state
appears inside the hybridization gap.
We also address the effect of the confining potential in optical lattice systems 
by means of real-space DMFT calculations.
}

\kword{periodic Anderson model, superfluid, dynamical mean-field theory,
optical lattice}

\begin{document}
\maketitle

\section{Introduction}
Ultracold atomic gases have  
attracted considerable interest~\cite{Review,PethickSmith,Pitaevskii}
since the successful realization of Bose-Einstein condensation 
in a bosonic $\rm ^{87}Rb$ system.\cite{Rb}
One of the most active topics in this field 
is the study of  fermionic optical lattice systems,
which are formed by loading ultracold fermions 
into a periodic potential~\cite{BlochGreiner,Bloch,Jaksch,Morsch}.
This setup provides a clean realization of a quantum lattice system, in which 
remarkable phenomena have been observed 
such as the Mott insulating state~\cite{Jordens,Schneider} 
and the BCS-BEC crossover 
in two component fermionic systems.~\cite{BCSBEC1,BCSBEC2,BCSBEC3}
Due to the high controllability of the lattice structure 
and onsite interactions, ultracold fermionic systems in optical lattices 
can be regarded as quantum simulators of the Hubbard model,
whose ground-state properties have been investigated 
by numerous theoretical and computational approaches. 
Recent papers have suggested the possible realization of
optical lattice systems described by other theoretical models such as 
multi-component models\cite{Ottenstein,Huckans,Fukuhara} and 
the Kondo lattice model\cite{Gorshkov,Feig,Feig2}.
This stimulates further investigations on lattice models
with strong correlations. 
Among them, the periodic Anderson model (PAM)
describing conduction and localized bands
is one of the most important models in condensed matter physics.
It captures the essential physics of some heavy-electron systems realized 
in rare-earth compounds.\cite{PAM}
An important point is that quantum critical behavior is expected 
in the PAM with both repulsive and attractive interactions,
and the model may thus provide a stage to discuss a quantum phase transition 
in an optical lattice system.
However, this type of instability, which may
be important to predict the low-temperature properties 
of optical lattice systems, has not been discussed 
in the PAM with attractive interactions.
Furthermore, the confining potential should play a crucial role
in understanding the quantum critical behavior in optical lattice systems
with attractive interactions.
Therefore, it is important to systematically investigate the  
low temperature properties in this model with localized and itinerant bands.

In this paper, we study the PAM with attractive interactions
by combining dynamical mean-field theory 
(DMFT)~\cite{Metzner,Muller,Georges,Pruschke} 
with the continuous-time quantum Monte
Carlo (CTQMC) approach.~\cite{Rubtsov}
First, we discuss how the superfluid state is realized 
on a uniform bipartite lattice
and determine the phase diagram.
By examining the dynamical properties in the superfluid state in detail,
we demonstrate that low energy 
in-gap states are induced in the density of states.
We also discuss the effect of the confining potential
by means of the real-space DMFT and investigate
how the Kondo insulating state spatially competes with 
the superfluid state.

The paper is organized as follows. 
In Sec. \ref{2}, we introduce the model Hamiltonian and explain
the particle-hole transformation in the PAM.
We briefly summarize our theoretical approach.
We demonstrate how the superfluid state competes with 
the Kondo insulating state at low temperatures
in Sec. \ref{3}.
The effect of the confining potential is discussed in Sec. \ref{4}.
A brief summary is given in the last section.

\section{Model and Method}\label{2}
We consider the periodic Anderson model with 
conduction and localized bands, which is described by the following
Hamiltonian,
\begin{eqnarray}
H&=&H_0+H_1,\label{eq1}\\
H_0&=&-t\sum_{\langle i,j \rangle \sigma} c_{i\sigma}^\dag c_{j\sigma}
+V\sum_{i\sigma}\left(c_{i\sigma}^\dag f_{i\sigma}
+f^\dag_{i\sigma} c_{i\sigma}\right)\nonumber\\
&&+E_f\sum_{i\sigma} n^f_{i\sigma} -\sum_{ia\sigma}\left(\mu_a+h_a \sigma\right) n^a_{i\sigma},\\
H_1 &=& -U\sum_i\left[n^f_{i\uparrow} n^f_{i\downarrow} -\frac{1}{2}\left(n^f_{i\uparrow}+n^f_{i\downarrow}\right)\right],
\end{eqnarray}
where $c_{i\sigma} (f_{i\sigma})$ annihilates a particle in the conduction 
(localized) band 
on the $i$th site with spin $\sigma$, and 
$n_{i\sigma}^a = a_{i\sigma}^\dag a_{i\sigma} \;\; (a = c$ and $f)$.
$t$ is the nearest-neighbor hopping matrix for the
conduction band, $V$ is the hybridization between the two bands, 
$U$ is the attractive interaction, and
$E_f$ is the energy level for the localized band.
$\mu_c (\mu_f)$ and $h_c (h_f)$ are 
the chemical potential and the magnetic field
for the conduction (localized) band.
In the following, we refer to the conduction and localized bands as
$c$ and $f$ bands, for simplicity.

First, we wish to note that the low-energy properties of 
the PAM with attractive interactions 
[eq. (\ref{eq1})] are closely related 
to those of the repulsive PAM. 
This can be understood by applying a particle-hole transformation 
to the model on the bipartite lattice.
It is explicitly given by 
\begin{eqnarray}
&\left\{
\begin{array}{l}
c_{i\sigma}\rightarrow\sigma\tilde{c}_{i\sigma}^\dag\\
f_{i\sigma}\rightarrow -\sigma\tilde{f}_{i\sigma}^\dag
\end{array}
\right.&i\in A,\\
&\left\{
\begin{array}{l}
c_{j\sigma}\rightarrow -\sigma\tilde{c}_{j\sigma}^\dag\\
f_{j\sigma}\rightarrow \sigma\tilde{f}_{j\sigma}^\dag
\end{array}
\right.&j\in B,
\end{eqnarray}
where $A$ and $B$ are sublattice indices, and 
$\tilde{c}_{i\sigma}$ ($\tilde{f}_{i\sigma}$) 
annihilates a particle in the $c$ ($f$) band 
on the $i$th site with spin $\sigma$.
We then obtain the PAM with repulsive interactions $\tilde{U}(=-U)$, where
the energy level for the $f$ band, 
chemical potentials and magnetic fields are transformed 
as $\tilde{E}_f = 0$, $\tilde{\mu}_f=h_f$, $\tilde{\mu}_c=h_c$, 
$\tilde{h}_f=E_f-\mu_f$, and $\tilde{h}_c=\mu_c$, respectively.
In particular, the low temperature properties of both models are equivalent 
at half filling $(E_f=0, \mu_a=0, h_a=0)$,
which is similar to the case of the Hubbard model.\cite{ph}

The low-temperature properties of the PAM with repulsive interactions,
which have been discussed in the context of $f$ electron systems,\cite{PAM}
may be helpful to understand those of our attractive model.
In the repulsive PAM,
the hybridization between the $c$ and $f$ bands favors the formation
of a local Kondo singlet state.
Therefore, a large hybridization stabilizes the Kondo insulating state 
at half filling and the heavy metallic state away from half filing.
\cite{Schweitzer,Mutou}
On the other hand,
the Coulomb repulsion favors magnetic correlations between sites,
which tends to induce an antiferromagnetically ordered state.
\cite{Dorin,JAP,Rozenberg,Imai,KogaNRG}
Therefore, in the half-filled PAM on a bipartite lattice, 
a quantum phase transition occurs between the Kondo insulating state
and the antiferromagnetically ordered state
in two and higher dimensions.
Note that this competition is controlled by
not only the interaction strength 
but also the magnetic field.
If a magnetic field is applied, 
the spin gap in the Kondo insulating state decreases
and finally vanishes.
At this point, a field-induced magnetic phase transition occurs, where
a spontaneous staggered magnetization appears 
in the plane perpendicular to the applied field.~\cite{Beach,Ohashi,Milat}

In the PAM with attractive interactions,
a similar competition is expected at low temperatures.
The corresponding ordered phase should be characterized by 
a pair potential $\Delta$, which is given by
\begin{eqnarray}
\Delta_a&=&\frac{1}{N}\sum_i \langle a_{i\uparrow}a_{i\downarrow}\rangle,
\end{eqnarray}
where $a = c, f$, and $N$ is the total number of sites.
This quantity corresponds to the staggered magnetization in the $x$ direction
$m_x^a=\frac{1}{N}\sum_i (-1)^i \langle \tilde{a}_{i\uparrow}^\dag\tilde{a}_{i\downarrow}
+\tilde{a}_{i\downarrow}^\dag\tilde{a}_{i\uparrow} \rangle$ 
in the PAM with repulsive interactions.
By contrast, it is known that the Kondo insulating state
has no order parameter since it is adiabatically connected 
to the paramagnetic state at high temperatures.
This implies that it is also realized 
in the PAM with attractive interactions.

To discuss how the superfluid state competes with the Kondo insulating state, we make use of DMFT. 
Since local particle correlations can be taken into account precisely,
this method is formally exact in the limit of infinite dimensions, 
and has successfully been applied to strongly correlated systems.
In DMFT, the lattice Green's function is obtained via a self-consistency 
condition imposed on the impurity problem.
The non-interacting Green's function for the lattice model is given as,
\begin{equation}
\hat{G}_0^{-1}(k, i\omega_n)=\left(
\begin{array}{cc}
\xi_c \sigma_0+\left(\mu_c-\epsilon_k\right)\sigma_z & -V \sigma_z \\
-V \sigma_z&\xi_f \sigma_0+\left(\mu_f-E_f\right)\sigma_z 
\end{array}
\right)
\end{equation}
where $\xi_a=i\omega_n +h_a\; (a=c,f)$, $\epsilon_k$ is the dispersion relation 
of the $c$ band,
$\sigma_z$ is the $z$ component of the Pauli matrix, 
$\sigma_0$ is the identity matrix, 
$\omega_n = (2n+1)\pi T$ is the Matsubara frequency, and 
$T$ is the temperature. 
Here, the Green's function is represented in the
Nambu formalism to describe the superfluid state.
The lattice Green's function is then given 
in terms of the site diagonal self-energy $\hat{\Sigma}(i\omega_n)$ as,
\begin{eqnarray}
{\hat G}(i\omega_n) &=&
\left(
\begin{array}{cccc}
G_c(i\omega_n)&G_{cf}(i\omega_n)\\
G_{fc}(i\omega_n)&G_f(i\omega_n)\\
\end{array}
\right)\\
&=&\int dk \left[{\hat G}_0^{-1}(k, i\omega_n) 
- {\hat\Sigma}(i\omega_n)\right]^{-1},\nonumber\\
{\hat \Sigma}(i\omega_n) &=& \left(
\begin{array}{cccc}
0&0\\
0&\Sigma_f(i\omega_n)\\
\end{array}
\right).
\end{eqnarray}
The self-energy for the $f$ band $\Sigma_f$ is
given as,
\begin{eqnarray}
\Sigma_f(i\omega_n) &=&\left(
\begin{array}{cc}
\Sigma_{f\uparrow}\left(i\omega_n\right) & S_f\left(i\omega_n\right)\\
S_f\left(i\omega_n\right) & -\Sigma_{f\downarrow}^*\left(i\omega_n\right)
\end{array}
\right),
\end{eqnarray}
where $\Sigma_{f\sigma}(i\omega_n) \; [S_f(i\omega_n)]$ 
is the normal (anomalous) part of the self-energy for the $f$ band. 
In DMFT, the self-consistency condition is given by 
\begin{eqnarray}
G_f\left(i\omega_n\right) &=& G_{imp}\left(i\omega_n\right),
\end{eqnarray}
where $G_{imp}$ is the Green's function 
of the effective impurity model.
The effective medium for each site is given by 
\begin{eqnarray}
{\cal G}^{-1}\left(i\omega_n\right) &=&\left[G_{f}
\left(i\omega_n\right)\right]^{-1}+
\Sigma\left(i\omega_n\right).
\end{eqnarray}

In the PAM with attractive interactions, a competition between 
the Kondo insulating state and the superfluid state is expected. 
Therefore, it is necessary that the impurity solver is capable of 
treating different energy  scales accurately.
The exact diagonalization method \cite{Caffarel} 
is efficient to
discuss ground state properties, but it is difficult
to quantitatively discuss the critical behavior and dynamical properties
at finite temperatures.
The numerical renormalization group method\cite{NRG,NRGreview} 
is one of the most powerful methods to 
describe the low energy states, but may not be appropriate for
investigating the higher energy region. 
In this paper, we use the recently developed CTQMC method.
In this approach, a Monte Carlo sampling of certain collections of diagrams is
performed in continuous time, and 
thereby the Trotter error, which originates from the Suzuki-Trotter 
decomposition, is avoided.
The CTQMC method comes in two flavors, a weak-coupling\cite{Rubtsov} and 
strong-coupling\cite{Werner} formalism, 
and is applicable to general classes of models such as the Hubbard model,
\cite{CTQMC,KondoCTQMC,MultiCTQMC,KogaWerner,Dao,KogaQMC2} 
the periodic Anderson model,\cite{Luitz} 
the Kondo lattice model, \cite{Otsuki}
and the Holstein-Hubbard model.\cite{Phonon}
Here, we use the continuous-time auxiliary field version of the weak-coupling 
CTQMC method \cite{Gull} extended to the Nambu formalism, 
which allows us to directly access the superfluid state 
at low temperatures.~\cite{KogaWerner}
In our CTQMC simulations, 
we measure normal and anomalous Green's functions 
on a grid of a thousand points. 
To discuss static and dynamical properties in the system quantitatively,
the number of CTQMC samplings in each DMFT iteration is changed, depending on 
the temperature, the chemical potential, and the magnitude of the hybridization:
{\it e.g.} we have performed eighty million (five billion) samplings 
for the half-filled system with $U=1$, $V=0.40$, and $T=0.02$ 
($U=1$, $V=0.18$ and $T=0.01$).
In the following, we set $\mu = \mu_c = \mu_f$ and $h=h_c=h_f=0$, for simplicity.

\section{Periodic Anderson model on the hypercubic lattice}\label{3}
\begin{figure}[htb]
\begin{center}
\includegraphics[width=7cm]{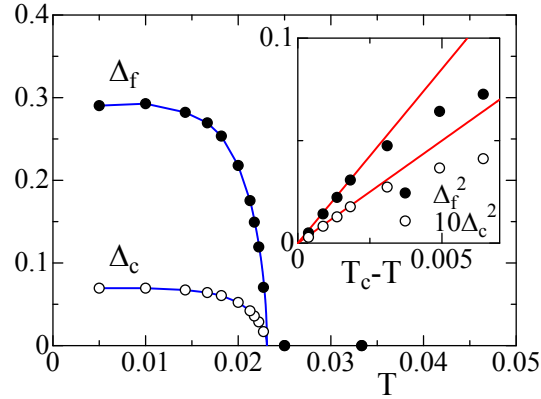}
\caption{(Color online) 
The pair potentials $\Delta_c$ and $\Delta_f$ 
for the superfluid state as a function of 
temperature $T$ when $U=1$ and $V=0.3$.
The inset shows the critical behavior of the pair potentials. 
Solid lines are guides to eyes.
}
\label{fig:dth}
\end{center}
\end{figure}

We first consider the PAM on the infinite-dimensional 
hypercubic lattice.
The bare density of states for the $c$ band is Gaussian with the bandwidth
$t^*$: $\rho(\omega) = \exp[-(\omega/t^*)^2]/\sqrt{\pi t^{*2}}$.
In this section, we use $t^*$ as the energy unit.
The normal state properties have been discussed 
in detail by means of DMFT, and the phase diagram has been obtained from 
the divergence of the staggered susceptibility 
at finite temperatures.\cite{JAP,Rozenberg,Imai,KogaNRG}
However, the off-diagonal ordered state has not been discussed directly
since it cannot easily be treated.
Here, we combine DMFT with the CTQMC method based on the Nambu formalism to
discuss how the superfluid state is realized at low temperatures.
In Fig.~\ref{fig:dth}, we show the temperature dependence of the pair potential
in the half-filled system with $U=1$ and $V=0.3$.
At high temperatures, the pair potential is zero and 
the normal state is realized.
As temperature is decreased below a critical value, the pair potentials for both bands are 
simultaneously induced,
which implies a phase transition to the superfluid state.
By examining the critical behavior $\Delta \sim |T-T_c|^\beta$ with 
the exponent $\beta=1/2$, 
we determine the critical temperature $T_c\sim 0.023$, 
as shown in the inset of Fig.~\ref{fig:dth}.

\begin{figure}[htb]
\begin{center}
\includegraphics[width=7cm]{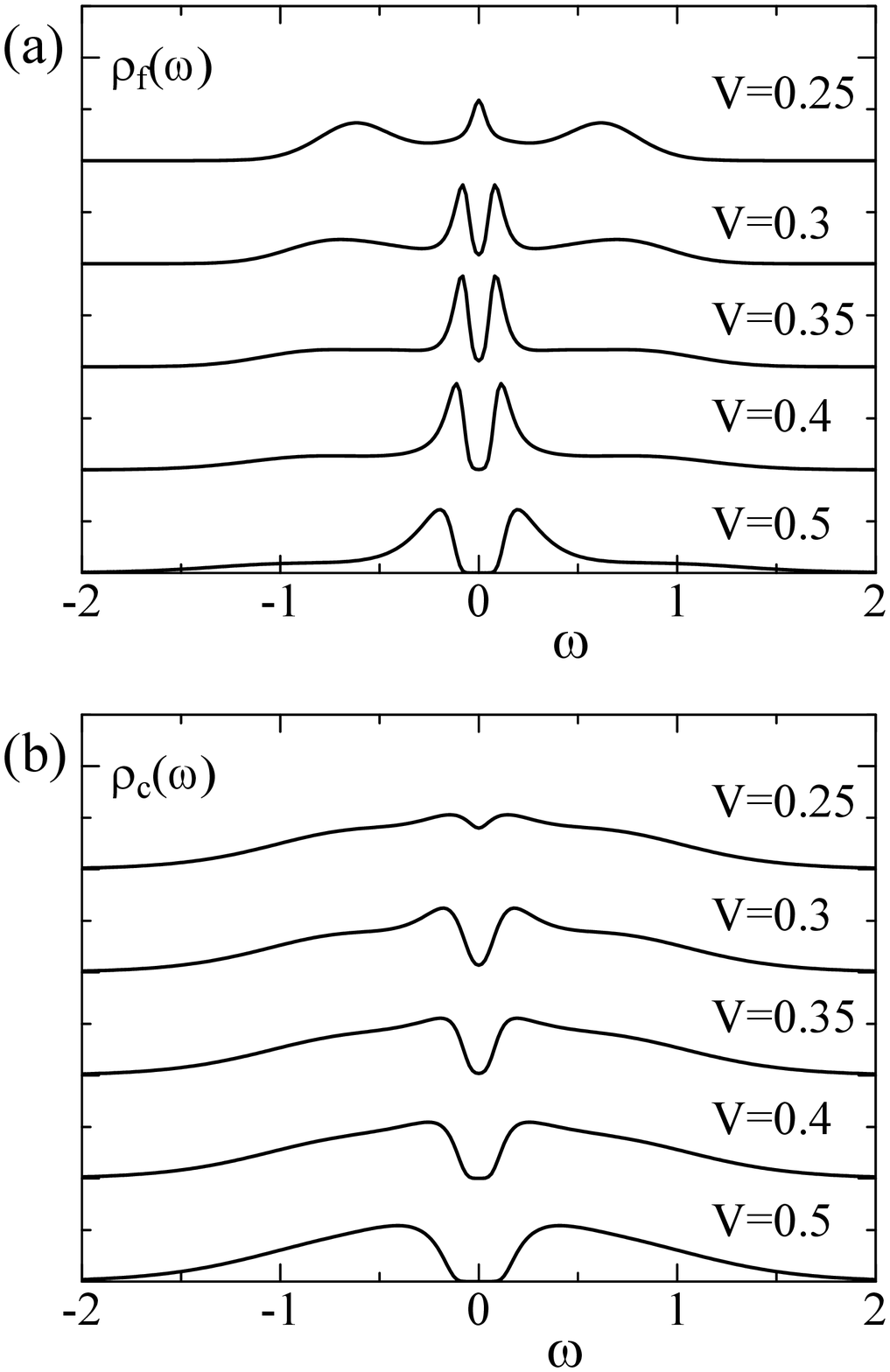}
\caption{(Color online) Density of states for localized and conduction bands 
when $V=0.25, 0.3, 0.35, 0.4$ and $0.5$ at $T=0.02$.}
\label{fig:dos-50}
\end{center}
\end{figure}

However, the superfluid state is not always realized 
at low temperatures in the PAM.
By performing similar calculations for several values 
of the hybridization $V$, 
\begin{figure}[htb]
\begin{center}
\includegraphics[width=7cm]{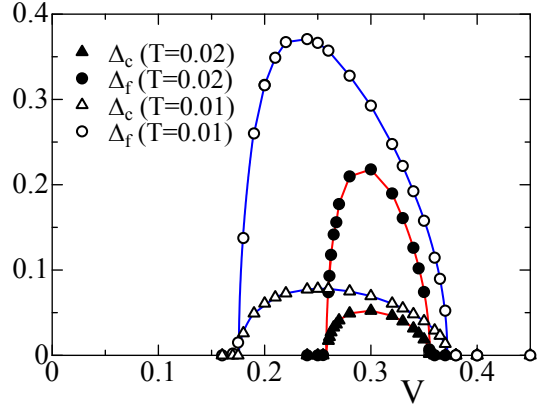}
\caption{(Color online) The pair potential as a function of the hybridization $V$ 
when $T=0.01$ (open symbols) and $T=0.02$ (solid symbols).
}
\label{fig:dv}
\end{center}
\end{figure}
we find, as shown in Fig.~\ref{fig:dv}, three distinct regions: 
$V<V_{c1}$, $V_{c1}<V<V_{c2}$ and $V_{c2}<V$.
The two critical points are obtained as
$V_{c1}\sim 0.26 (0.18)$ and $V_{c2} \sim 0.36 (0.37)$ at $T=0.02 (0.01)$.
To clarify the nature of these states, 
we show the density of states obtained by the maximum entropy method 
in Fig.~\ref{fig:dos-50}.
When $V>V_{c2}$, 
the pair potential is zero and a charge gap appears around the Fermi level 
in each band.
This implies that the large hybridization stabilizes the Kondo insulating state
at half filling.
In the intermediate region $(V_{c1}<V<V_{c2})$,
the superfluid state is realized since the system has finite pair potentials
and the corresponding gap in the density of states.
Note that no drastic change in the density of states appears at
$V=V_{c2}$, in contrast to the pair potential.
A similar behavior has been  
discussed in the half-filled Kondo lattice model.\cite{Bodensiek}
On the other hand, 
when $V<V_{c1}$,
a finite density of states appears around the Fermi level in each band.
This implies that a normal metallic state is realized.

\begin{figure}[htb]
\begin{center}
\includegraphics[width=7cm]{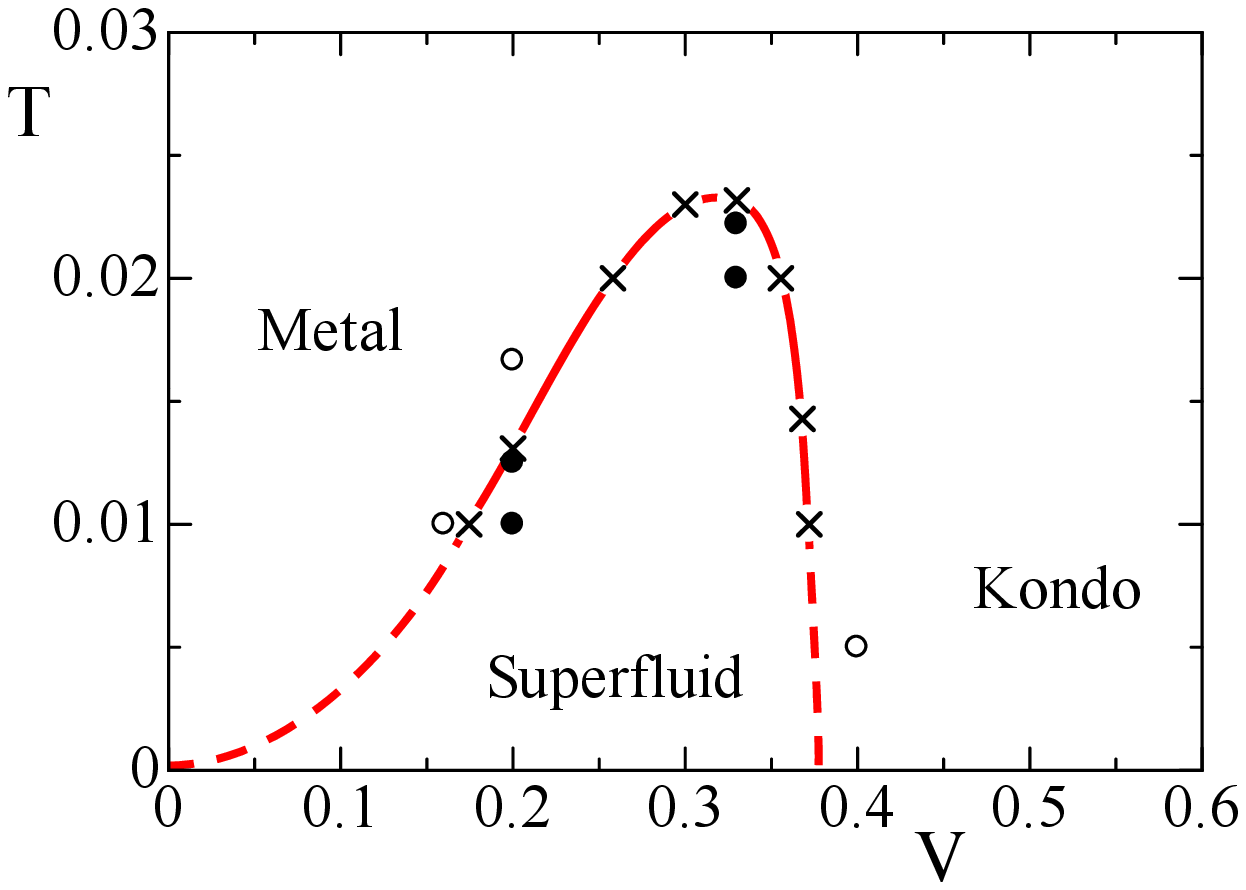}
\caption{(Color online) Phase diagram of the half-filled PAM with $U=1$.
Crosses are the critical points, solid (open) circles indicate
the state with (without) the pair potential. 
The phase boundary is a guide to the eyes.}
\label{fig:PDH}
\end{center}
\end{figure}

By performing similar calculations, we obtain the phase diagram 
for the half filled system with $U=1$, 
as shown in Fig.~\ref{fig:PDH}.
The normal state is realized at higher temperatures, while
a different behavior appears at lower temperatures, 
depending on the magnitude of the hybridization.
In the small $V$ case,
the system is reduced to the non-interacting $c$ band weakly coupled 
to the correlated $f$ band, which makes
the critical temperature for the superfluid state rather low.
Therefore, the normal metallic region shrinks as temperature is decreased, 
$V_{c1}\rightarrow 0\; (T\rightarrow 0)$, as shown in Fig.~\ref{fig:PDH}.
On the other hand, we find that 
the other critical point $V_{c2}$ between the superfluid state and 
the Kondo insulating state is slightly increased as temperature is lowered.
We expect that at zero temperature, the superfluid state is realized
in the small $V$ region and a quantum phase transition 
occurs to the Kondo insulating state at $V_{c2} (\sim 0.38)$.
This is consistent with ground-state properties for 
the PAM with repulsive interactions\cite{Rozenberg}
and the Kondo lattice model,\cite{KondoCTQMC,Otsuki}
where the Kondo insulator competes with 
the magnetically ordered state.

\begin{figure}[htb]
\begin{center}
\includegraphics[width=12cm]{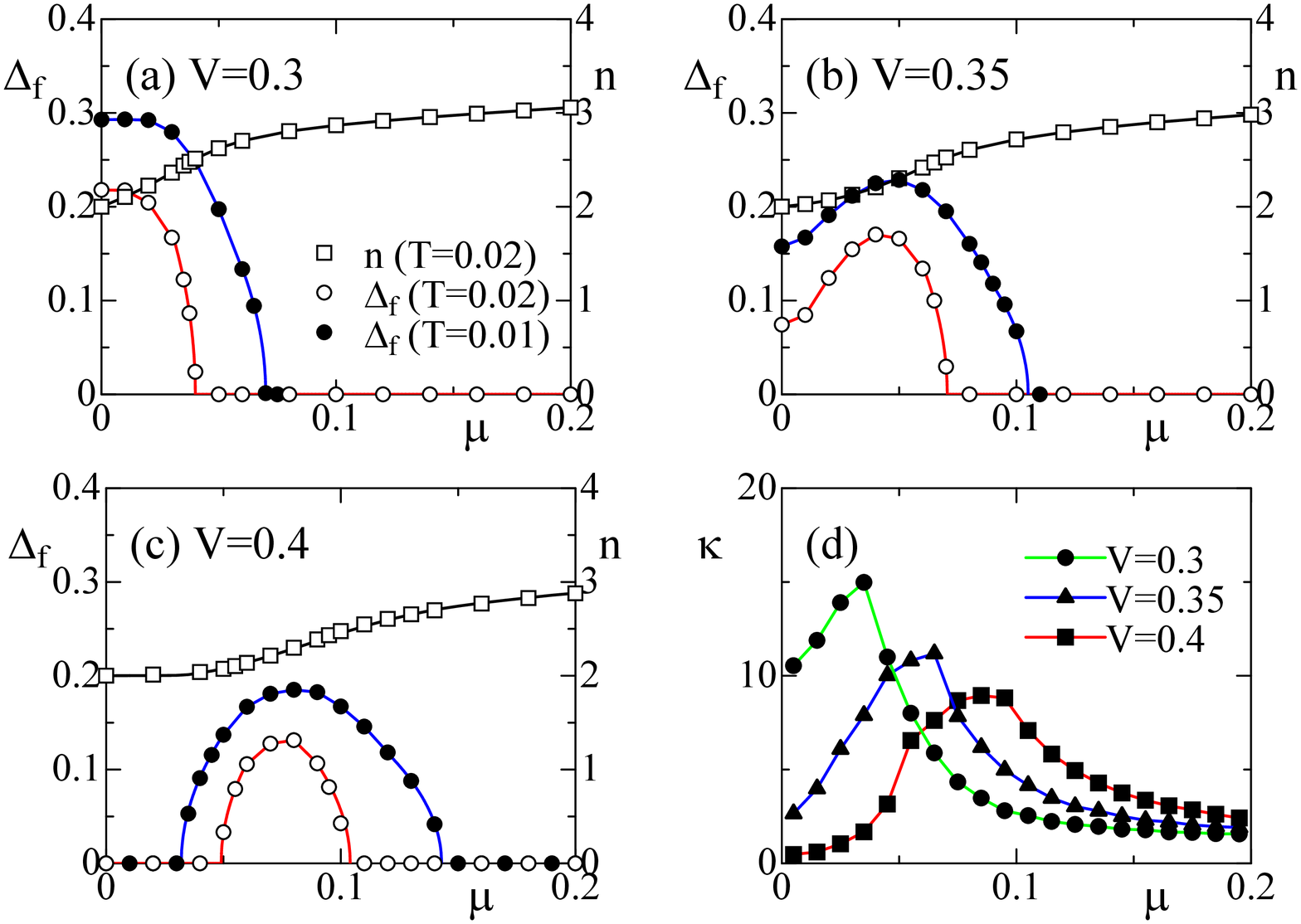}
\caption{(Color online) The local particle density and 
the pair potential for the $f$ band 
as a function of the chemical potential 
when $V=0.3$ (a), $V=0.35$ (b), and $V=0.4$ (c).
(d) The compressibility of the system with $V=0.3, 0.35$ and $0.4$ at $T=0.02$.
}
\label{fig:mu}
\end{center}
\end{figure}

Now, let us consider how the Kondo insulating state competes with
the superfluid state away from half filling.
In the following, we show only the pair potential $\Delta_f$ since
$\Delta_c$ behaves essentially the same.
In Fig.~\ref{fig:mu}, we show the local particle density at each site
$n (=\sum_{ai\sigma} \langle n_{i\sigma}^a\rangle /N)$
and the pair potential $\Delta_f$ when the chemical potential is varied.
Here, we show only the local particle density 
at $T=0.02$ in Figs.~\ref{fig:mu} (a), (b) and (c) 
since $n(T=0.01)$ is almost identical.
In addition, the compressibility $\kappa (=\partial n/\partial\mu)$ 
deduced from the numerical derivative is also shown in Fig.~\ref{fig:mu}~(d).
This quantity corresponds to the magnetic susceptibility in the $z$ direction
in the repulsive PAM. 
Therefore, a cusp singularity is naively expected at the critical point.
When $V=0.3$, the half-filled system $(\mu=0)$ is in the superfluid state 
at low temperatures, as discussed above.
Shifting the chemical potential away from zero,  
the pair potential is little affected and 
the local particle density varies smoothly, as shown in Fig.~\ref{fig:mu}~(a).
Therefore, we conclude that the superfluid state is stable 
against small changes in the chemical potential.
Far away from half filling,
the pair potential decreases and finally reaches zero. 
A phase transition then occurs to the normal metallic state. 
By contrast, a different behavior appears in the case $V=0.35$.
The pair potential initially increases as the chemical potential 
is shifted away from zero,
has a maximum value around $\mu \sim 0.05$, and finally
vanishes at the phase transition, as shown in Fig.~\ref{fig:mu}~(b).
The nonmonotonic behavior is even more clearly evident in the case $V=0.4$,
as shown in Fig.~\ref{fig:mu}~(c).
When $\mu<\mu_{c1}$, where $\mu_{c1}\sim 0.049 (0.032)$ at $T=0.02 (0.01)$, 
the pair potential is zero and
the local particle density is little changed.
In fact, Fig.~\ref{fig:mu}~(d) shows the strong suppression in 
the compressibility.
This implies that an incompressible Kondo insulating state 
is realized at half filling, instead of the superfluid state.
An increase in the chemical potential beyond $\mu_{c1}$ 
drives a chemical potential-induced superfluid phase transition,  
marked by the appearance of the pair potential and 
a cusp singularity in the compressibility
although the latter is not clearly visible in Fig.~\ref{fig:mu}~(d). 
Further increase in $\mu$ changes the local particle density and 
finally drives the system
to the normal metallic state at $\mu=\mu_{c2}$, where 
$\mu_{c2}\sim 0.10 (0.14)$ at $T=0.02 (0.01)$.

\begin{figure}[htb]
\begin{center}
\includegraphics[width=7cm]{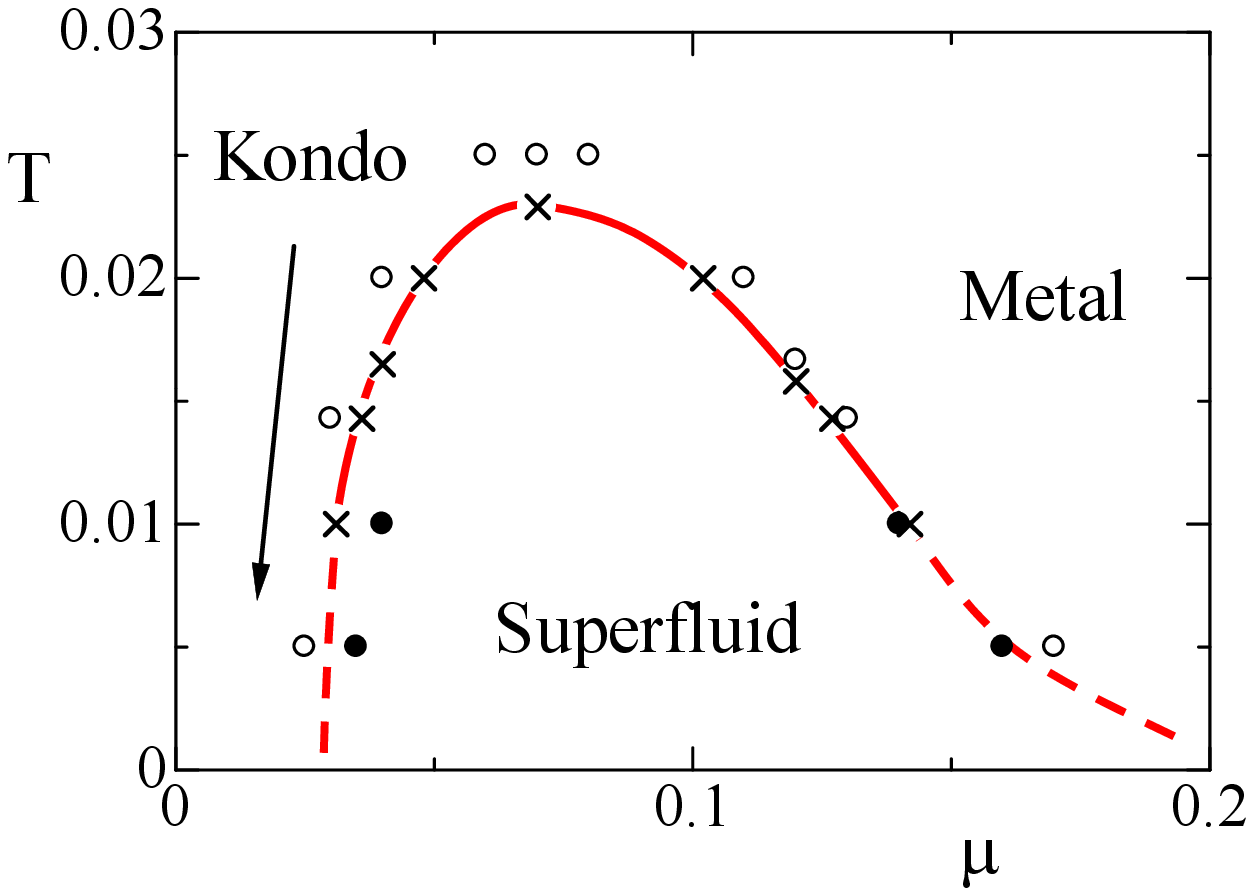}
\caption{(Color online) Phase diagram for the system with $U=1$ and $V=0.4$.
Crosses are the critical points, solid (open) circles indicate
the state with (without) the pair potential.  
The phase boundary is a guide to the eyes.}
\label{fig:phase4}
\end{center}
\end{figure}

\begin{figure}[htb]
\begin{center}
\includegraphics[width=12cm]{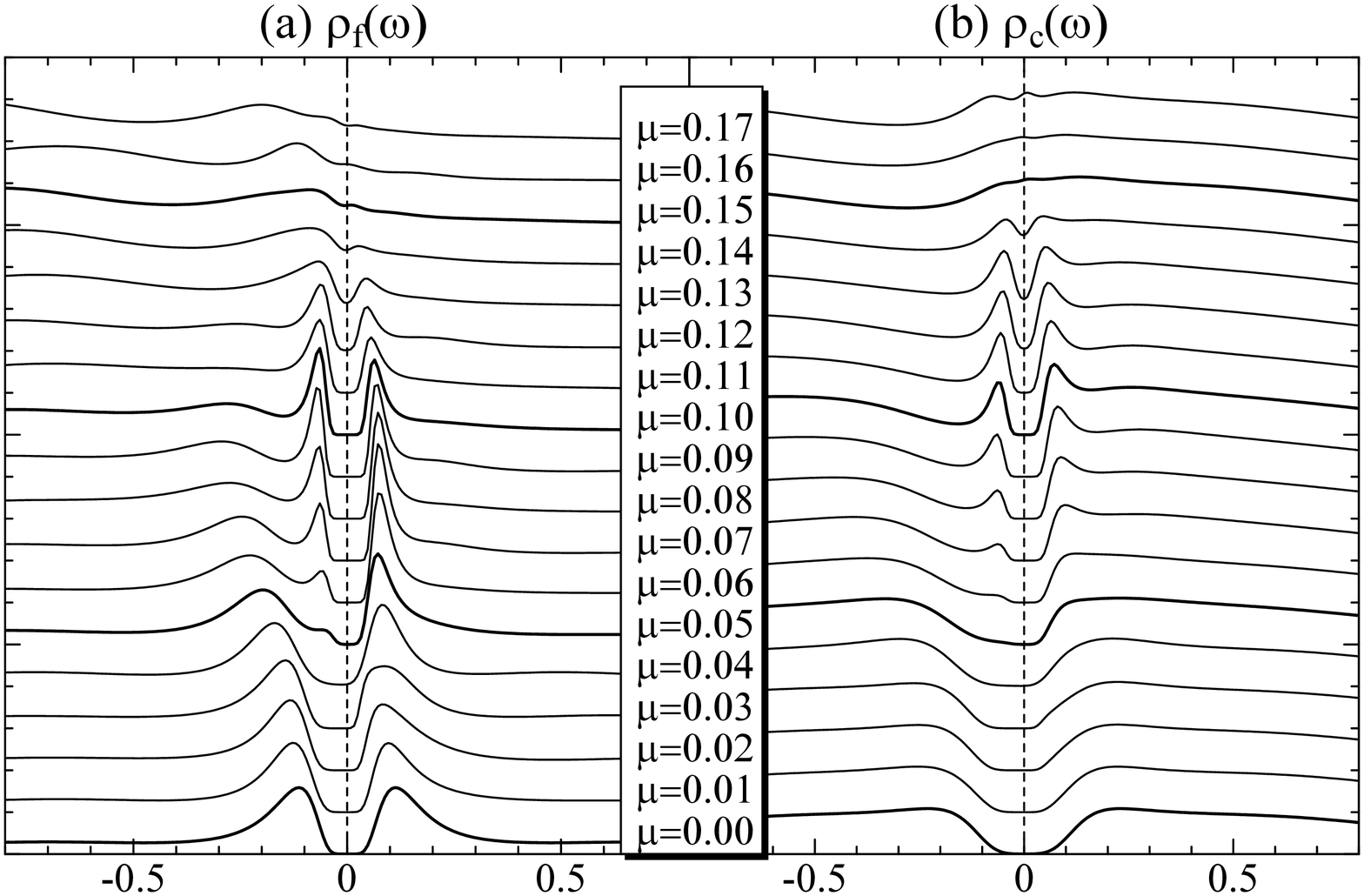}
\caption{Density of states for the $f$ (a) and $c$ (b) bands 
of the system with $U=1$ and $V=0.4$ at $T=0.01$.
The data are for the chemical potentials $\mu=0.00, 0.01, \cdots, 0.17$
from the bottom to the top.}
\label{fig:dos}
\end{center}
\end{figure}

By performing similar calculations for various temperatures,
we obtain the phase diagram for $U=1$ and $V=0.4$ shown
in Fig.~\ref{fig:phase4}.
This is essentially the same as the phase diagram for the magnetic field
in the PAM with repulsive interactions.\cite{Beach,Milat,Ohashi}
When the system is half filled $(\mu=0)$,
the Kondo insulating state is realized at low temperatures.
At $T=0.01$,
two peak structures clearly appear at the edges of the hybridization gap
in the density of states of the $f$ band, 
as shown in Fig.~\ref{fig:dos}~(a).
The introduction of the chemical potential simply shifts these peaks 
in the case $\mu < 0.03$ since the Kondo state is incompressible.
Further increase leads to interesting behavior 
in the density of states. 
We find that around $\mu\sim 0.05$, the upper peak never approaches 
the Fermi level and low energy states are, instead, induced 
below the Fermi level between the two peaks. 
This implies that particles in the vicinity of the Fermi level form 
Cooper pairs, which yields low-energy in-gap states.
The latter interpretation is consistent with the fact that 
the chemical potential-induced superfluid phase transition occurs 
at $\mu_{c1}\sim0.032$.
When the chemical potential increases beyond the critical point, 
the low-energy feature grows and becomes the dominant peak.
It is also found that as the chemical potential is varied,
the center of the gap is always located at the Fermi level,
in contrast to the case with $\mu < \mu_{c1}$.
On the other hand, 
the peak at the lower edge of the hybridization gap becomes smeared.
Therefore, we conclude that a compressible superfluid state 
is stabilized in this region.
When the system approaches $\mu_{c2}(\sim 0.14)$, 
the superfluid gap collapses and
another phase transition occurs to the normal metallic state.
A similar behavior is also found in the density of states
of the $c$ band, as shown in Fig.~\ref{fig:dos}~(b).
We emphasize that our calculation directly treats the low temperature superfluid 
state, in contrast to previous works for the PAM.\cite{Beach,Milat,Ohashi}
Figure~\ref{fig:dos} therefore clarifies 
how the chemical potential-induced phase transition 
yields low energy in-gap states in the density of states.

The results obtained above may have important consequences for three-dimensional 
optical lattice systems.
In particular, one might expect quantum critical behavior 
in the vicinity of a certain surface 
in the three-dimensional optical lattice system if
the confining potential is regarded as a site-dependent 
chemical potential.
In the following section, 
we study the effect of the confining potential carefully to discuss
the competition between the Kondo insulating state and the superfluid state
in such optical lattice systems.

\section{Effect of the confining potential}\label{4}

We study here the effect of the confining potential in the PAM.
As discussed in the previous section, we have found that
the Kondo insulating state competes with the superfluid state 
at low temperatures.
Furthermore, it has been clarified that in a certain parameter region,
a shift of the chemical potential can  
stabilize the superfluid state.
Therefore, if the confining potential is simply
regarded as a site-dependent chemical potential,
the Kondo insulating state and the superfluid state may be separated by 
a quantum critical surface in the three-dimensional optical lattice system.
In this case, the critical temperature 
should depend on the position in the trap. 
However, the physics in the presence of a trap is 
highly nontrivial and therefore 
it is necessary to discuss the low temperature properties 
using more sophisticated tools.

\begin{figure}[htb]
\begin{center}
\includegraphics[width=7cm]{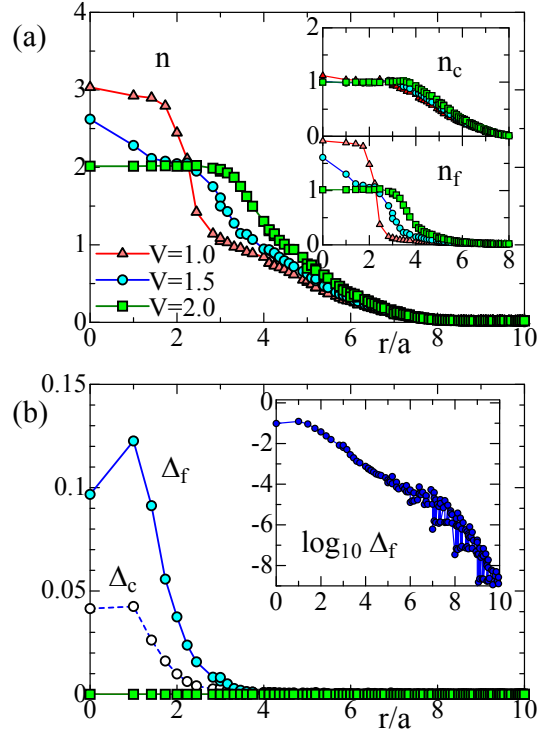}
\caption{(Color online) Local particle density~(a) and 
the pair potentials~(b) as a function of the distance from 
the center in the three-dimensional optical lattice system. 
Triangles, circles, and squares represent the results in the system with 
$V=1.0, 1.5$ and $V=2.0$ when $U=3.5$ and $T=0.033$.
The inset of figure~(a) shows 
the local particle density for the $c$ and $f$ bands.
The pair potential is shown on a logarithmic scale in the inset of 
the figure~(b).
}
\label{fig:nr}
\end{center}
\end{figure}

To this end, we use the real-space DMFT.
The method is based on the local approximation, but
intersite correlations are to some extent taken into account.
In fact, the real-space DMFT approach has been successfully used 
to discuss low temperature properties 
of optical lattice systems\cite{Helmes,KogaQMC2,MAGRDMFT,Koga} and
the interface between transition metal oxides.\cite{Okamoto}
Here, making use of this method, 
we consider a three-dimensional optical lattice system.
The Hamiltonian for the confining potential is defined as,
\begin{eqnarray}
H_c &=& \sum_{ia\sigma} v(r_i) n_{i\sigma}^a,
\end{eqnarray}
where $v(r) = v_0 (r/a)^2$, $v_0$ is the curvature of the confining potential,
$r_i$ is the distance between the $i$th site and the center of the trap, and
$a$ is the lattice constant.
In this section, we fix $t=1$ and $v_0=0.1$.
We then treat a system with 4169 sites restricted by 
the condition $(r/a \le 10)$.
Using the CTQMC method as an impurity solver, 
we perform the real-space DMFT calculations.
We obtain results for 
the spatial distributions of the local particle density and 
the pair potentials for  
the system with $U=3.5$ and $\mu=0.5$,
as shown in Fig.~\ref{fig:nr}.
When the hybridization is small $(V=1.0)$, 
the localized $f$ particles are weakly coupled 
to the $c$ band.
Therefore, the $f$ band is almost occupied in the core 
$r/a\lesssim\sqrt{\mu/v_0}\sim 2.2$, and empty outside the core
due to the attractive interaction for the $f$ band.
By contrast, the noninteracting $c$ particles are widely distributed 
due to the hopping integral between sites,
as shown in the inset of Fig.~\ref{fig:nr}~(a).
In this case, no pair potential appears and the metallic state is realized.
An increase in the hybridization affects the distribution of the $f$ band
and low-temperature properties.
When $V=1.5$, 
the local particle density around the center is decreased and 
the pair potential is induced in the system, 
as shown in Fig.~\ref{fig:nr}.
Therefore, the superfluid state is realized 
in the whole region with $n(r)\neq 0$ although
the pair potential is tiny away from the center, 
as shown in the inset of Fig.~\ref{fig:nr}~(b). 
This behavior is similar to that in the attractive Hubbard model 
with a confining potential.\cite{KogaQMC2}
Note that 
the rapid decay in the pair potential sets in around $r/a=1$,
in contrast to the distribution of the local particle density.
This may be explained by the competition between the Kondo insulating state 
and the superfluid state.
We find a shoulder structure around $r/a=2$ 
in the local particle density,
which indicates a tendency to form the Kondo insulating state 
in the superfluid state.
This results in the rapid decay of the pair potential.
Further increase in the hybridization tends to result in a plateau structure
and decreases the pair potential. 
When $V=2.0$, the pair potential vanishes and the plateau clearly appears 
at a value of $n(r)$ corresponding to half of the states occupied.
Therefore, the Kondo insulating state is stabilized 
in the core region with $r/a\lesssim 3$.

With these calculations, 
we have clarified that the Kondo insulating state spatially competes with
the superfluid state in an optical lattice system with confining potential.
However, we could not find a critical surface 
between the superfluid state and the Kondo insulating state, 
which could naively be expected from the results of the infinite dimensional PAM.
Therefore, we can say that 
intersite correlations play an important role and must be considered in a discussion  
of the low temperature properties of optical lattice systems.

Before concluding the paper,
we would like to comment on correlation effects in the $c$ band,
which can be tuned by the Feshbach resonance in optical lattice systems.
According to previous studies of the PAM with repulsive interactions,
these interactions do not affect the low temperature properties qualitatively.
\cite{Schork,Sato}
Therefore, it is expected that in the attractive case, 
the essence of the low temperature properties must be 
described by the simple PAM [eq. (\ref{eq1})]
although phase boundaries may be slightly shifted by 
interactions between the $c$ particles.

\section{Summary}
We have investigated the periodic Anderson model with attractive interactions
on the hypercubic lattice.
By combining DMFT with the CTQMC method based on the Nambu formalism,
we have studied quantitatively 
how the superfluid state is stabilized at low temperatures.
It has been found that 
a low-energy state characteristic of the superfluid phase 
appears in the hybridization gap near the critical chemical potential.
We have also discussed the effect of the confining potential 
in the three-dimensional optical lattice by means of the real-space DMFT
to clarify how the superfluid state spatially competes 
with the Kondo insulating state.

\section*{Acknowledgement}

This work was partly supported by the Grant-in-Aid for Scientific Research 
20740194 (A.K.) and 
the Global COE Program ``Nanoscience and Quantum Physics" from 
the Ministry of Education, Culture, Sports, Science and Technology (MEXT) 
of Japan. PW acknowledges support from SNF Grant PP002-118866.

\end{document}